\documentstyle[eqsecnum,aps,epsfig]{revtex}
\begin{document}
\draft
\title{The low-energy nuclear density of states and the saddle point 
approximation}
\author{Sanjay K. Ghosh\footnote{Email : sanjay@triumf.ca} and 
Byron K. Jennings\footnote{Email : jennings@triumf.ca}} 
\address{TRIUMF, 4004 Wesbrook Mall, Vancouver, British Cloumbia, Canada 
V6T 2A3}
\date{\today}
\maketitle
\begin{abstract}
The nuclear density of states plays an important role in nuclear reactions.
At high energies, above a few MeV, the nuclear density of states is well 
described
by a formula that depends on the smooth single particle density of states at
the Fermi surface, the nuclear shell correction and the pairing energy. 
In this
paper we present an analysis of the low energy behaviour of the nuclear 
density of states using the saddle point approximation and extensions to it. 
Furthermore, we prescribe a
simple parabolic form for excitation energy, in the low energy limit,
which may facilitate an easy computation of level densities.

\end{abstract}
\pacs{21.10.-k, 21.10.Ma, 26.50.+x}

\section{Introduction}

One of the important ingredients in the Hauser-Feshbach approach to the 
calculation
of nuclear reaction rates important for astrophysical interest is the nuclear
density of states\cite{rausher}. In fact uncertainties in the nuclear density
of states is a leading cause of errors\cite{rausher} in these calculations. 

The studies of nuclear level densities dates back to the 1950s. with work by
Rozenweig\cite{rosenweig}, and Gilbert and Cameron\cite{cameron,gilbert}. The usual
technique is to calculate the partition function and then invert the Laplace
transform using the saddle point approximation. At energies sufficiently high
for shell and pairing effects to be washed out the density of states is given
in terms of the single particle density of states at the Fermi surface (and
its derivatives) the shell correction energy and pairing energy. Most
statistical model calculations use the back shifted fermi gas description
\cite{gilbert}. Monte-Carlo shell model calculations\cite{dean} as well as
combinatorial approaches\cite{paar} show excellent agreement with this 
phenomenological approach.  At lower energies
the results are more problematic and typically crude extrapolations from the
higher energy are used. 

In this paper we study the nuclear level density with an emphasis on
 the lower energy region, using a single particle shell model.
The dependences in the two regimes are rather different. 
 In contrast to higher energies, where the density of states depends 
on the shell correction and the smooth single particle density of states, in 
the lower energy
regime the density of states depends on the separation of single particle 
levels and their degeneracy. 
Moreover, at very low energy the saddle point approximation itself breaks 
down. We show that in this region the correction suggested by Grossjean 
and Feldmeier\cite{grossjean} gives dramatic improvements. 

In the next section we review the saddle point approximation in the context
of nuclear
level density calculation. We use thermodynamic identities to rewrite the
equations in simpler form compared to the usual ones\cite{huizenga}. 
Furthermore, we discuss the possible ways to simplify the evaluation of 
level densities at low energies. A temperature dependent parabolic equation
for excitation energy seems to be a good choice. The corrections 
suggested\cite{grossjean} and the corresponding modifications to the equation 
are discussed in section 3. Finally in section 4 we discuss our calculation 
and the results. 

\section{The Saddle Point Approximation}

The grand canonical partition function for two type of particles can be 
written as:
\begin{equation}
\label{oone}
e^{\Omega }=\sum _{N',Z',E'}\exp (\alpha _{N}N'+\alpha _{Z}Z'-\beta E')
\end{equation}
 where the sum is over all nuclei with \( N' \) neutrons, \( Z' \) protons
and over all energy eigenstates \( E' \) . \( \tau = \beta^{-1} \),
is the temperature and \( \mu_{N(Z)} = \frac {\alpha_{N(Z)}}{\beta} \) is
the chemical potential for neutrons (protons). The sum over eigenstates can 
be substituted by an integral:
\begin{equation}
\label{otwo}
e^{\Omega }=\sum _{N',Z'}\int dE'\rho (E',N',Z')\exp (\alpha _{N}N'
+\alpha _{Z}Z'-\beta E')
\end{equation}
 where \( \rho (E',N',Z') \) is the nuclear density of states. It represents
the density of energy eigenvalues for the nucleus \( (N', Z') \) at the
energy \( E' \). The above equation also shows that the grand partition 
function can be considered a Laplace transform of the nuclear density
of states. The inversion
integral is:
\begin{equation}
\label{finalro1}
\rho (E',N',Z')=\frac{1}{(2\pi i)^{3}}\oint d\alpha _{N}\oint d\alpha _{Z}\oint \beta e^{S}
\end{equation}
where the entropy \( S=\Omega -\alpha _{N}N-\alpha _{Z}Z+\beta E \) . 
The above
contour integrals are also known as Darwin- Fowler integral. This integral is
usually done by the saddle point approximation and in this section we will 
explore this approximation. The location of the saddle point is defined by 
the equations,
\begin{equation}
\frac{d\Omega '}{d\beta }=-E\rm {;}\hspace {1cm}
\frac{d\Omega '}{d\alpha _{N}}=N\rm 
{;}\hspace {1cm}\frac{d\Omega '}{d\alpha _{Z}}=Z.
\end{equation}

The path of the integration can be chosen to pass through this point. By
expanding the exponent \( S \) in Taylor series about the saddle point and
retaining only the quadratic terms, the nuclear density of states in the
saddle point approximation can be written as:
\begin{equation}
\label{rhotwo}
\rho =\frac{e^{S}}{(2\pi )^{3/2}D^{1/2}}
\end{equation}
where \( D \) is the determinant of the second derivative of \( S \) with
respect to the parameters \( \alpha _{N} \) , \( \alpha _{Z} \) , and 
\( \beta . \)
The determinant can be simplified to a product of factors by changing the
variables which are held fixed when the derivatives are performed. 

The determinant is written as: 

\begin{equation}
\label{denone}
D=\left| \begin{array}{ccc}
\frac{d^{2}S}{d\beta ^{2}} & \frac{d^{2}S}{d\beta d\alpha _{N}} & \frac{d^{2}S}{d\beta d\alpha _{Z}}\\
\frac{d^{2}S}{d\beta d\alpha _{N}} & \frac{d^{2}S}{d\alpha _{N}^{2}} & \frac{d^{2}S}{d\alpha _{N}d\alpha _{Z}}\\
\frac{d^{2}S}{d\beta d\alpha _{Z}} & \frac{d^{2}S}{d\alpha _{N}d\alpha _{Z}} & \frac{d^{2}S}{d\alpha ^{2}_{Z}}
\end{array}\right| 
\end{equation}
To simplify the above determinant  we change the independent variables to 
\( \tau =1/\beta  \)
, \( \mu _{N}=\tau \alpha _{N} \) , and \( \mu _{N}=\tau \alpha _{Z} \) and
change the dependent variable to \( \Omega '=\tau \Omega =\tau S+\mu _{N}N
+\mu _{Z}Z-E \).
In terms of the new variables the equations determining the saddle point are:

\begin{equation}
\frac{d\Omega '}{d\tau }=-S\rm {;}\hspace {1cm}\frac{d\Omega '}{d\mu _{N}}=-N\rm 
{;}\hspace {1cm}\frac{d\Omega '}{d\mu _{Z}}=-Z.
\end{equation}
 Using these equations the determinant can be written as: 
\begin{equation}
\label{det1}
D=-\tau ^{5}\left| \begin{array}{ccc}
\frac{dS}{d\tau } & \frac{dN}{d\tau } & \frac{dZ}{d\tau }\\
\frac{dS}{d\mu _{N}} & \frac{dN}{d\mu _{N}} & \frac{dZ}{d\mu _{N}}\\
\frac{dS}{d\mu _{Z}} & \frac{dN}{d\mu _{Z}} & \frac{dZ}{d\mu _{Z}}
\end{array}\right| 
\end{equation}
In deriving this result we have used the fact that in a determinant addition 
of a multiple of row
(column) to another row (column) does not change the value of the
determinant. In the first row the derivatives
are at constant \( \mu _{N} \) and \( \mu _{Z} \) ; in the second at constant
\( \tau  \) and \( \mu _{Z} \), and the third at constant \( \tau  \) and
\( \mu _{N} \) . The variables which are held constant can be changed using 
the following equations:
\begin{equation}
\left. \frac{dS}{d\tau }\right| _{\mu _{N}\mu _{Z}}=\left. 
\frac{dS}{d\tau }\right| _{NZ}+\left. \frac{dS}{dN}\right| _{\tau Z}\left. 
\frac{dN}{d\tau }\right| _{\mu _{N}\mu _{Z}}+
\left. \frac{dS}{dZ}\right| _{\tau N}\left. 
\frac{dZ}{d\tau }\right| _{\mu _{N}\mu _{Z}}
\end{equation}
\begin{equation}
\left. \frac{dS}{d\mu _{N}}\right| _{\tau \mu _{Z}}=\left. 
\frac{dS}{dN}\right| _{\tau Z}\left. \frac{dN}{d\mu _{N}}\right| _{\tau \mu _{Z}}+
\left. \frac{dS}{dZ}\right| _{\tau N}\left. 
\frac{dZ}{d\mu _{N}}\right| _{\tau \mu _{Z}}
\end{equation}
and
\begin{equation}
\left. \frac{dS}{d\mu _{Z}}\right| _{\tau \mu _{N}}=\left. 
\frac{dS}{dN}\right| _{\tau Z}\left. 
\frac{dN}{d\mu _{Z}}\right| _{\tau \mu _{N}}+
\left. \frac{dS}{dZ}\right| _{\tau N}
\left. \frac{dZ}{d\mu _{Z}}\right| _{\tau \mu _{N}}
\end{equation}

Subtracting \( \left. \frac{dS}{dN}\right| _{\tau Z} \) times the second 
column and \( \left. \frac{dS}{dZ}\right| _{\tau N} \) times the third 
column from the first column we have:
\begin{equation}
D=-\tau ^{5}\left. \frac{dS}{d\tau }\right| _{NZ}\left| \begin{array}{cc}
\frac{dN}{d\mu _{N}} & \frac{dN}{d\mu _{Z}}\\
\frac{dZ}{d\mu _{N}} & \frac{dZ}{d\mu _{Z}}
\end{array}\right| 
\end{equation}

This procedure can be repeated to yield:
\begin{equation}
\label{finaldet1}
D=-\tau ^{5}\left. \frac{dS}{d\tau }\right| _{NZ}
\left. \frac{dN}{d\mu _{N}}\right| _{\tau Z}
\left. \frac{dZ}{d\mu _{Z}}\right| _{\tau \mu _{N}}
\end{equation}
Note the progression on the variables held constant. This procedure can be extended
in an iterative manner to any number of constants of motions. The main gain
is that we now have a simple product rather than a determinant. 

Let us now try to understand how this new form may help us in practice. 
To calculate the density of states as
a function of energy for fixed particle number we need the entropy \( S \)
as a function of energy at fixed \( N \) and \( Z \) for the exponent in the
numerator. The temperature can then be obtained from \( \left. dS/dE\right| _{NZ}=
1/\tau  \)
and \( \left. dS/d\tau \right| _{NZ}=-1/(\tau ^{3}d^{2}S/dE^{2}) \). This leaves
the derivatives of the particle numbers to be separately evaluated. Thus we
have three independent functions to parameterize. 

To see that this modified form of the density of states agrees with the standard
form we consider the independent particle model where with a constants single
particle density of states \( g \). The entropy then is \( S=2\sqrt{{aE}} \)
where \( a=\pi g/6. \) The temperature is \( \tau =\sqrt{{E/a}} \), 
\( \left. dS/d\tau \right| _{NZ}=a/2 \),
and \( dN/d\mu _{N}=dN/d\mu _{Z}=g \). This then gives the well know formulae:

\[
\rho =\frac{\sqrt{\pi }}{12}\frac{\exp [2\sqrt{aE}]}{E^{5/4}a^{1/4}}\]
 Now let us consider a normal quantum system with a discrete spectrum. In this
case there are in general no closed forms for the various functions, so we consider
the zero temperature limit of \( dN/d\mu _{N} \). We start with an open shell
situation where there is a partially filled shell. For this discussion it is
only necessary to consider the properties of the partially filled level. We
take the level to have a degeneracy of \( g_{1} \), an energy of 
\( \epsilon _{1} \)
and a partial occupancy of \( d \). As \( \tau  \) goes to zero the saddle
point condition for the number of particle becomes 
\( g_{1}/(1+\exp [(\epsilon _{1}-\mu )/\tau ])=dg_{1} \).
Solving for \( \mu  \) we have \( \mu =\epsilon _{1}+\tau \ln [d/(1-d)] \).
Note that this formulae breaks down for \( d \) equals zero or one corresponding
to closed shells. The derivative \( dN/d\mu  \) is given as \( g_{1}d(1-d)/\tau  \).
Note that it diverges as \( \tau  \) goes to zero. 

For a closed shell it necessary to consider two levels, the last filled level
and the first unfilled level. We denote the energies and degeneracies of these
levels as \( \epsilon _{1},g_{1} \) and \( \epsilon _{2},g_{2} \). The saddle
point condition is now \( g_{2}/(1+\exp [(\epsilon _{1}-\mu )/\tau ])+g_{2}/(1+\exp [(\epsilon _{2}-\mu )/\tau ])=g_{1} \).
Solving for \( \mu  \) we have \( \mu =(\epsilon _{1}+\epsilon _{2})/2-\tau \ln [g_{1}/g_{2}] \)
for small temperatures. The derivative \( dN/d\mu  \) is given as \( 2\sqrt{g_{1}g_{2}}\exp [(\epsilon _{1}-\epsilon _{2})/(2\tau )]/\tau  \).
This goes to zero exponentially fast as \( \tau  \) goes to zero. Note that in
neither of the cases above shell correction is involved.

The above discussion is useful since  $S$ is a function 
of the energy. Here again one may use a few trick. It turns out to be easier 
to
parameterize $E$ as function of $\tau$. Since $\tau dS= dE$, one can write 
$S = \int_0^\tau \frac{1}{\tau'} \frac{dE}{d\tau'} d\tau' +
S(\tau=0)$. 
The last term is the integration constant and is given
once the degeneracy of the ground state is known. It contributes to
exponent but not to the denominator where derivatives are taken. 

Next one needs $E$ as a function of $\tau$. For many systems there is a
quite reliable approximation. For very low temperature, much less then
the level spacing, the energy does not change significantly. However
above some critical temperature ,$\tau_0$, it starts to increase
rapidly. For temperature nears this region the energy can be
parametrized quite simply by 
\begin{equation}
\label{fit}
E-E_0 = c (\tau-\tau_0)^2 \theta(\tau-\tau_0) 
\end{equation}
We have checked this approximation using a simple
shell model and found that it works quite well except if there are
more then one level approximately equal distant from the Fermi
surface. The parameters $\tau_0$ and $c$ depend on the level spacing
and degeneracy near the Fermi surface. Again they do not depend on the
shell correction. 

Before being useful at very low energies a short-coming of the saddle
point approximation must be overcome. It is well known that at low
energies the saddle point approximation tends to diverge as the
denominator goes to zero, In many cases this problem can be fixed by
using a technique from ref.~\cite{grossjean} which handles the
contribution to the nuclear density of states from the ground state
delta function explicitly. 

\section{Modified Saddle Point}
In ref.\cite{grossjean} Grossjean and Feldmeier have proposed a modification of 
the saddle point method to remove the divergences of the level density at the 
ground state. Introducing explicitly the ground state energy $E_g(A)$ as the
lower boundary, the density of state becomes
\begin{equation}
\tilde{\omega}(E^*, A) = \omega(E^{*}+E_{g}(A), A) - \delta(E^*)\delta(A-A_0)
\end{equation}
where $E^* = E - E_g$ is the excitation energy and $A_0$ is the mean particle 
number. The corresponding modified grand canonical potential is given by,
\begin{equation}
\label{tildeom}
\tilde{\Omega} = \Omega + \beta E_g + ln(1-Y)
\end{equation}
where 
\begin{equation}
\label{defy}
Y=d_{0} e^{\alpha_{N}N + \alpha_{Z}Z - \beta E_{g} - \Omega}
\end{equation}
The chemical potentials for neutrons and protons are given by 
$\mu_{Z}=\frac{\alpha_{Z}}{\beta}$ and $\mu_{N}=\frac{\alpha_{N}}{\beta}$
respectively, $d_{0}$ being the ground state occupancy.

The nuclear level density in the modified saddle point approximation becomes,
\begin{equation}
\label{finalro2}
\rho = \frac{\tilde{S}}{(2\pi)^{3/2} \tilde{D}^{1/2}}
\end{equation}
where $\tilde{D}$ is the determinant in the form of eq.\ref{det1}, 
defined in terms of $\tilde{S}$, the entropy corresponding to the potential
$\tilde{\Omega}$. The steps of the previous section 
(eq.(\ref{det1}) - eq.(\ref{finaldet1})) can be retraced and then one gets,
\begin{equation}
\label{finaldet2}
\tilde{D}=-\tau ^{5}\left. \tilde{\frac{dS}{d\tau }}\right| _{NZ}\left. 
\tilde{\frac{dN}{d\mu _{N}}}\right| _{\tau Z}\left. 
\tilde{\frac{dZ}{d\mu _{Z}}}\right| _{\tau \mu _{N}}
\end{equation}
The derivation of eq.(\ref{finaldet2}) is straightforward as it depends only
on the thermodynamic relations of the quantities involved and not on their
explicit forms.

The computation of the level density using eq.(\ref{finalro2}) will depend on the
relations between the usual and the modified thermodynamic quantities as the
usual quantities are directly related to the single particle shell model
states.

The modified saddle point conditions in terms of the usual thermodynamic 
potential becomes,
\begin{eqnarray}
\left. \frac{d\Omega}{d\alpha}\right| _{\alpha_{0}, \beta_{0}} = A
  \nonumber \\
\left. \frac{d\Omega}{d\beta}\right| _{\alpha_{0}, \beta_{0}} = -(E_{g} + 
\tilde{E^*})
\end{eqnarray}
where $\tilde{E^*}=E^{*} (1 - Y)$ (see eq.(\ref{defy})). For \( \alpha =
\alpha_{N(Z)} \), \( A = N(Z) \). The derivatives of the 
entropy and numbers are related as,
\begin{eqnarray}
\left. \tilde{\frac{dS}{d\tau}}\right| _{N,Z}  =
\left. (\frac{dS}{d\tau}\right| _{N,Z}  - \beta^{3}{E^*}^{2}Y +
\beta^{3}{E^*}^{2}Y^2) \times \frac{1}{-1 + Y}   \nonumber \\
\left. \tilde{\frac{d{N}}{d\mu_{N}}}\right| _{\tau,Z}  =
\left. \frac{d{N}}{d\mu_{N}}\right| _{\tau,Z} \times \frac{1}{-1 + Y} 
\nonumber \\
\left. \tilde{\frac{dZ}{d\mu_{Z}}}\right| _{\tau,N}  =
\left. \frac{dZ}{d\mu_{Z}}\right| _{\tau,N}  \times \frac{1}{-1 + Y} 
\end{eqnarray}

Using the above equations one can compute the level density 
(eq.(\ref{finalro2})) in terms of single particle states.

\section{discussion}
The single particle energies, required for the evaluation of different
thermodynamic quantities are obtained for the Nilsson shell 
model. The values of the constants associated with
\( {\bf{l}}^2 \) and \( {\bf{l.s}} \) are taken from ref.\cite{bohr}. 
Here it should be mentioned
that the level densities are strongly dependent on the single particle
energy levels. Hence for astrophysical applications one should make a 
judicious choice for the model as well as the constants.

We first calculate the level densities for different nuclei, for both the 
usual as well as modified saddle point approximations,  using all the filled
and a equal number of unfilled levels. It should be noted the for low 
excitation energies (of the order of first excitation level or less) 
the inclusion of only 
the last filled and the first filled level, as described in section 2, is 
sufficient for the evaluation of  the level densities.

A comparison of the level densities from eq.(\ref{finalro1}) 
and eq.(\ref{finalro2}) for nuclei $^{32}S$, $^{88}Sr$ and $^{208}Pb$ 
are shown in fig. 1, fig. 2 and fig. 3 respectively. The modified saddle
point results are shown by curve (a) and the usual saddle points results are
shown by curve (c) in the above figures. As evident from the 
graphs the usual saddle point does show a divergence at the low energies 
whereas modified version goes to zero smoothly. This is due to the fact that 
entropy in the modified version goes to zero much faster as it takes 
into account
the nonavailability of states below the ground state energy. Moreover, at 
low energies the differences are more
pronounced for the lighter nuclei like $^{32}S$ compared to $^{208}Pb$. The
differences can be attributed to the fact that in modified prescription
the thermodynamic potential gets an additive contribution compared to
the usual one as shown by eq.(\ref{tildeom}).  

As discussed in section 2. we try to fit the excitation energies with 
parabolic form as given in eq.(\ref{fit}). These fits for different nuclei 
are shown in fig. 4, fig. 5 and fig. 6 respectively. In figures (4-6) we
have plotted the variation of excitation energy with temperature. For low
energies, the fitted value is in good agreement with exact values. 
Next we calculate the entropy and its derivative, using the steps given 
in section 2, from this fitted expression for 
excitation energy, the derivatives of \( N \) and \( Z \) being the same as
in preceding paragraph. Using these we calculate the level densities for
different nuclei. A comparison of the level densities from the full 
calculation from modified saddle point approximation (curve (a)) and the 
one using fitted excitation energies (curve (b)) are shown in fig. 1, 
fig. 2 and fig. 3. It is
obvious from the graphs that the fitted excitation energy gives a better
agreement with exact calculation for heavier nuclei. 

To conclude, we have shown that the modification of the saddle approximation 
is necessary for the correct evaluation of the level densities at lower
energies. One can simplify the equations substantially using the 
thermodynamic identities. Furthermore, a parabolic prescription for the
excitation energy may be useful for easier computation of the level 
densities. More work in this direction is needed to make the methodology 
useful for direct application to astrophysical reactions.

\begin{figure}
\epsfig{file=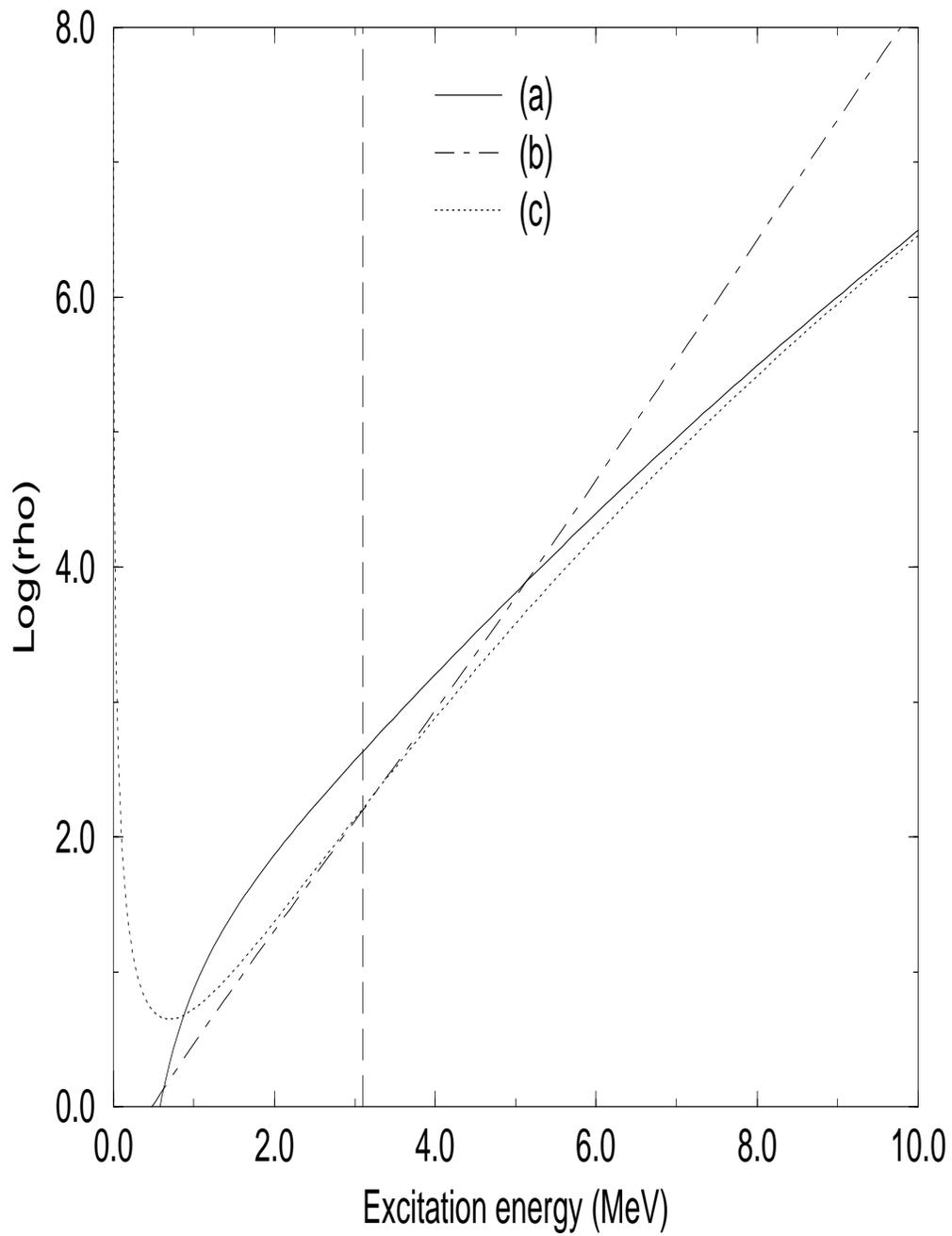,width=6.0in,height=8.0in}
{\centering
\caption{Level density for ${^{32}}S$; (a) modified saddle point, 
(b) corresponds to the fitted excitation energy as in 
fig. 4 and (c) usual saddle point. The vertical dashed line gives the 
position of 1st excitation level}} 
\end{figure}

\newpage

\begin{figure}
\epsfig{file=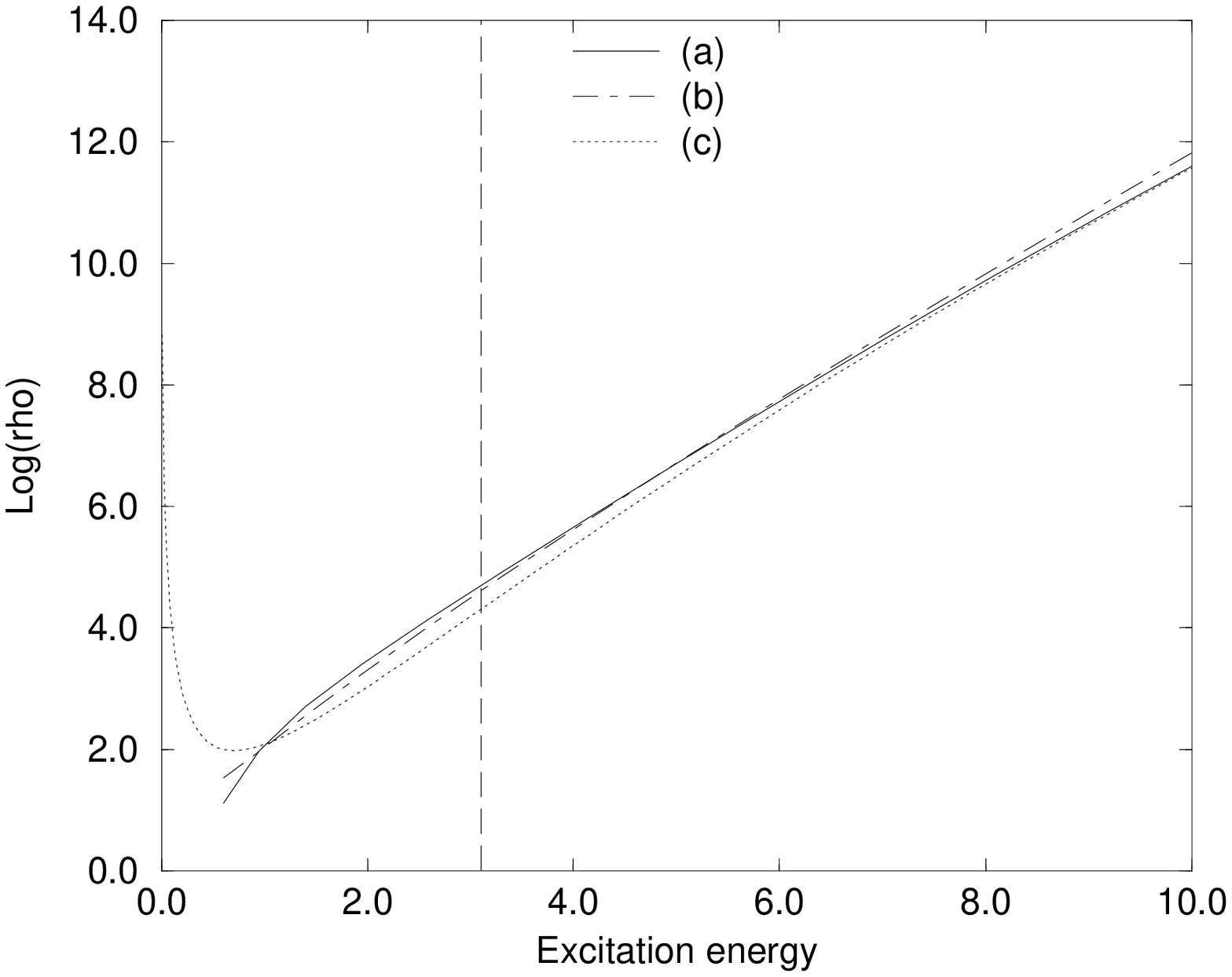,width=6.0in,height=8.0in}
{\centering
\caption{Level density for ${^{88}}Sr$; (a) modified saddle point, 
(b) corresponds to the fitted excitation energy as in 
fig. 4 and (c) usual saddle point. The vertical dashed line gives the 
position of 1st excitation level}} 
\end{figure}

\newpage
\begin{figure}
\epsfig{file=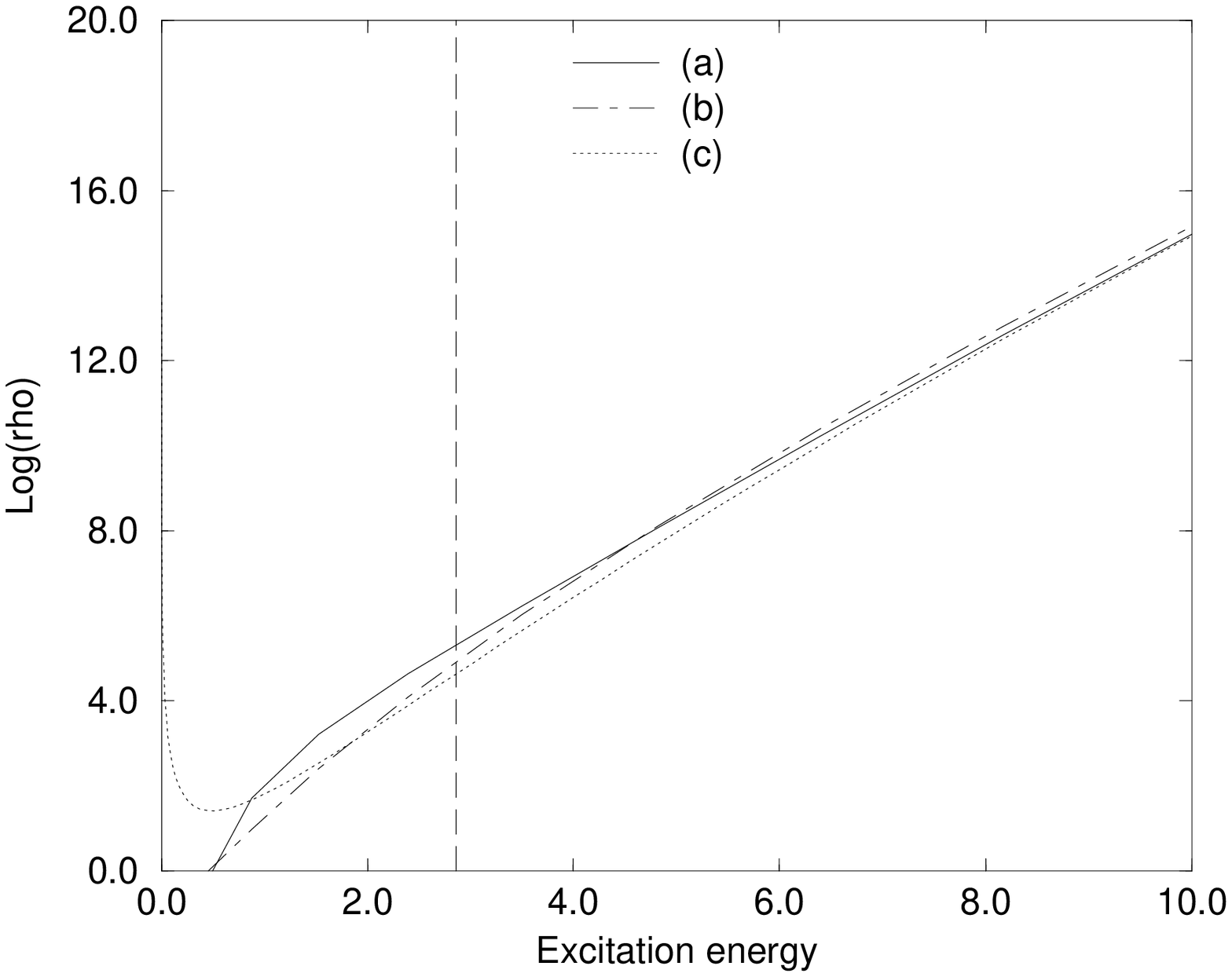,width=6in, height=8in}
{\centering
\caption{Level density for ${^{208}}Pb$; (a) modified saddle point, 
(b) corresponds to the fitted excitation energy as in 
fig. 4 and (c) usual saddle point. The vertical dashed line gives the 
position of 1st excitation level}} 
\end{figure}

\newpage

\begin{figure}
\epsfig{file=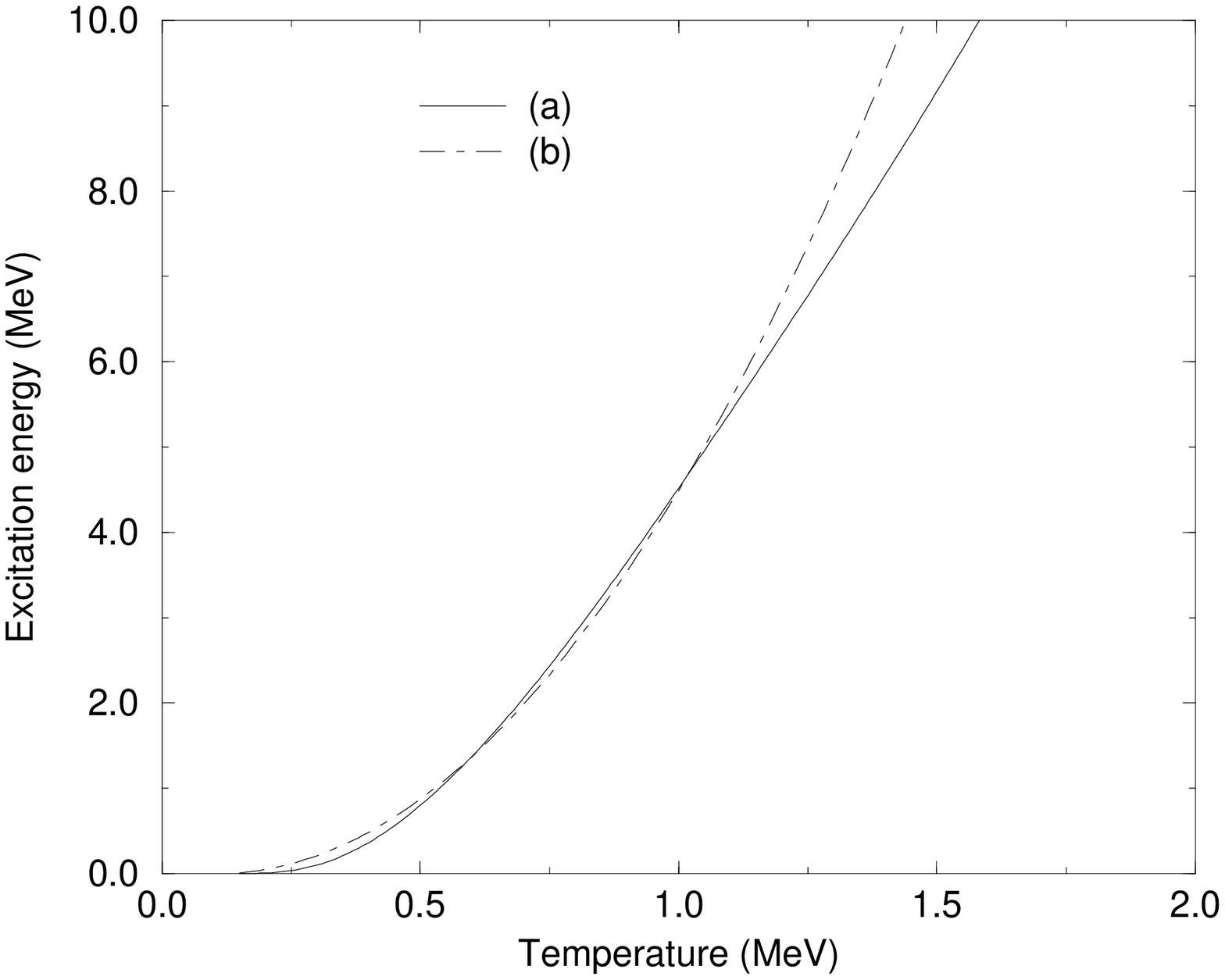,width=6in, height=8in}
{\centering
\caption{Excitation energy for ${^{32}}S$; (a) actual excitation energy nd
(b) corresponds to the fitted form eq.(\ref{fit}) with $E_{0}$=5.62 and 
$\tau_{0}$=0.11 }}
\end{figure}

\newpage
\begin{figure}
\epsfig{file=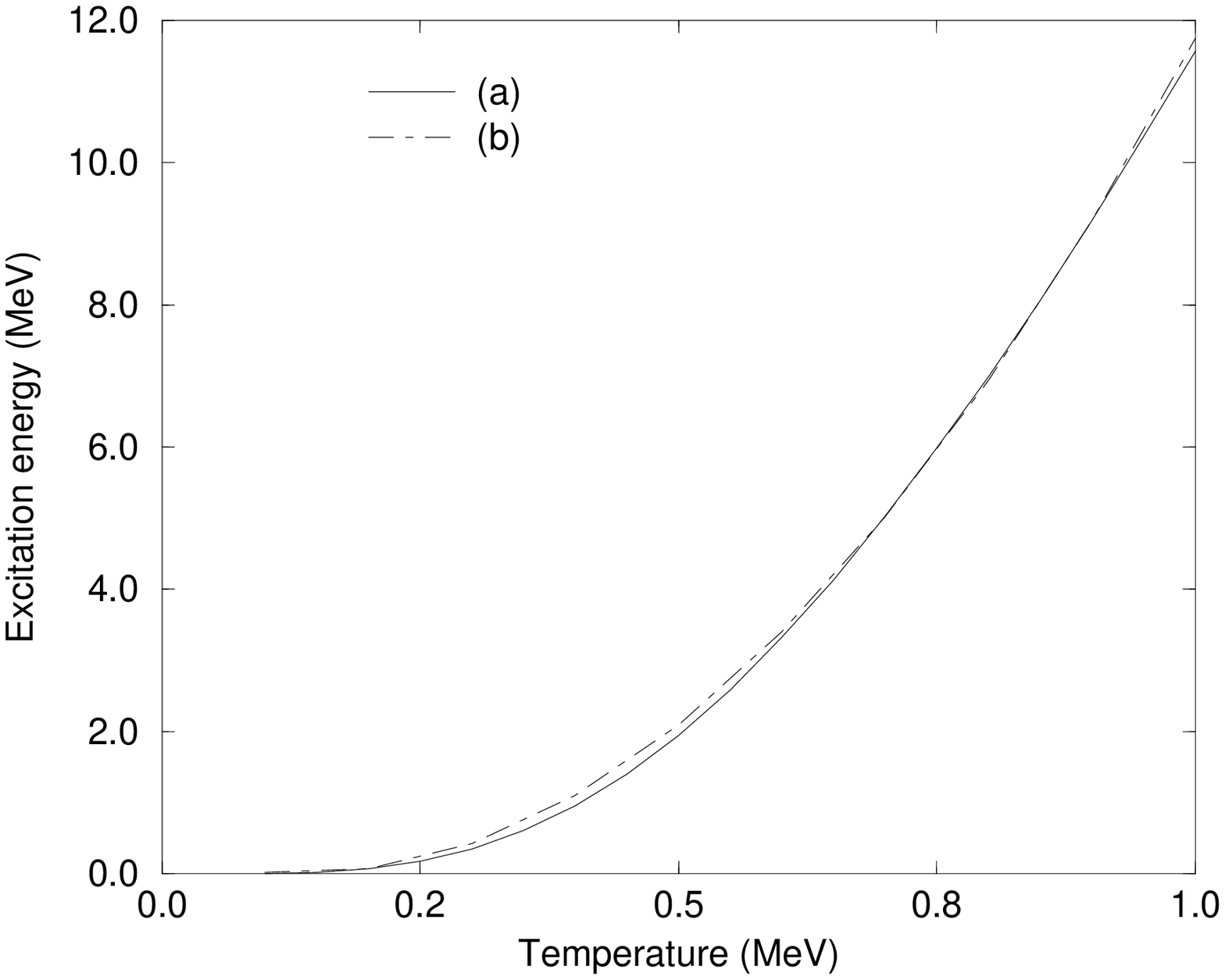,width=6in, height=8in}
{\centering
\caption{Excitation energy for ${^{88}}Sr$; (a) actual excitation energy nd
(b) corresponds to the fitted form eq.(\ref{fit}) with $E_{0}$=15.71 and 
$\tau_{0}$=0.14 }}
\end{figure}

\newpage
\begin{figure}
\epsfig{file=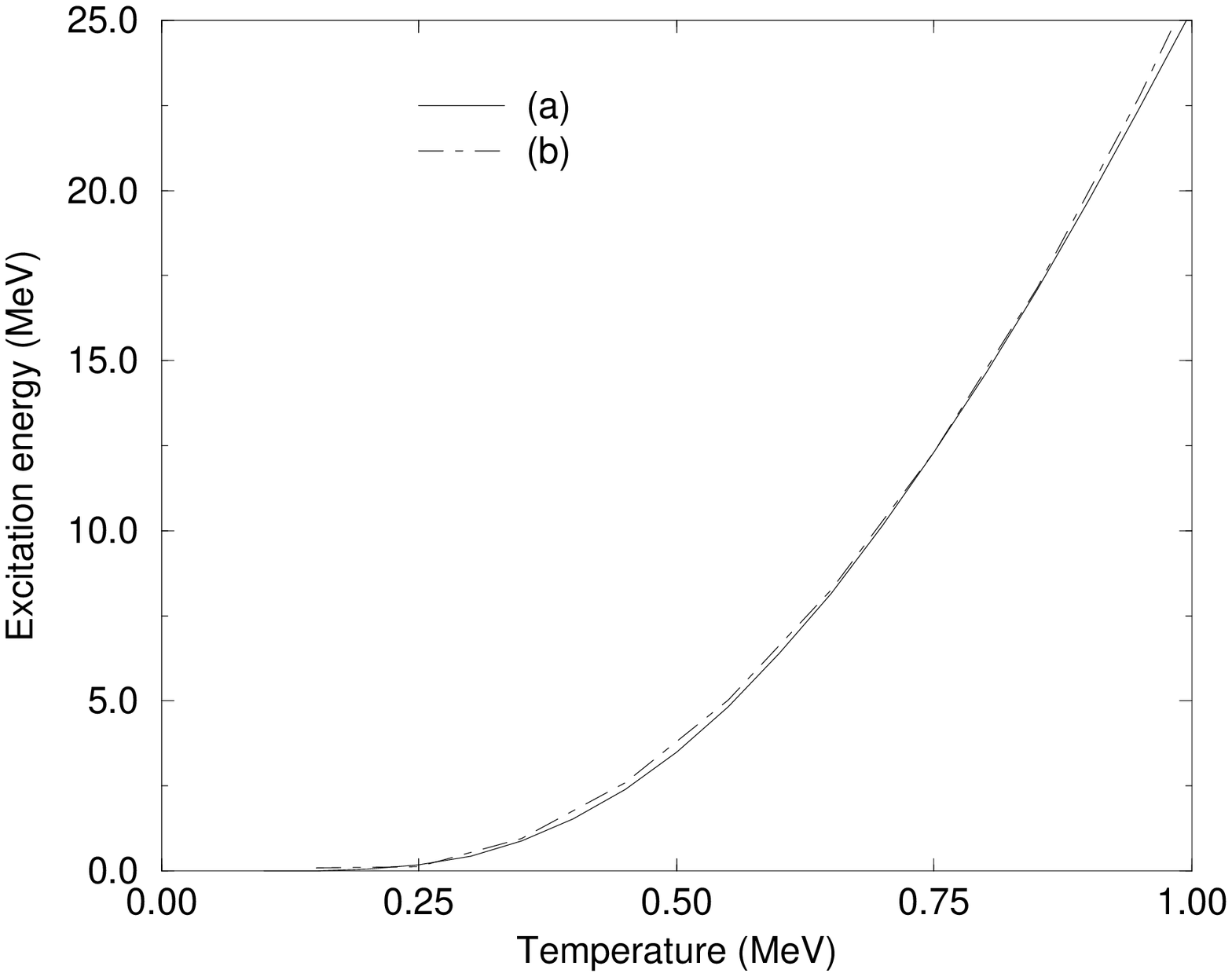,width=6in, height=8in}
{\centering
\caption{Excitation energy for ${^{208}}Pb$; (a) actual excitation energy nd
(b) corresponds to the fitted form eq.(\ref{fit}) with $E_{0}$=40.0 and 
$\tau_{0}$=0.20 }}
\end{figure}

\begin{references}
\bibitem{rausher}T.~Rausher, F.-K.~Theielemann and K.-L.~Kratz, Phys. Rev. 
{\bf C56}, 1613 (1997).
\bibitem{rosenweig}N,~Rosenweig, Phys.~Rev.~{\bf 105}, 950 (1957); {\bf 108}, 817
(1957).
\bibitem{cameron}A.G.~Cameron, Can.~J.~Phys. {\bf 36}, 1040 (1958).
\bibitem{gilbert} A.~Gilbert and A.G.~Cameron, {\bf 43}, 1446 (1965).
\bibitem{dean} D. J. Dean, S. E. Kroonin, K. -H. Langnake, P. B. Radha and
Y. Alhassid, Phys. Rev. Lett. {\bf 74}, 2909 (1995).
\bibitem{paar} V. Paar and R. Pezer, Phys. Rev. {\bf C55}, R1637 (1997).
\bibitem{grossjean}M.K.~Grossjean and H.~Feldmeier, Nucl.~Phys.~{\bf A444}, 
113 (1985).
\bibitem{huizenga}J. R. Huizenga and L. G. Moretto, Ann. Rev. Nucl. Sc.
{\bf 22}, 427 (1972).
\bibitem{bohr} A. Bohr and B. R. Mottelson, Nuclear Structure Vol. II,
World Scientific, (1998).

\end{references}
\end{document}